\documentclass{article}
\usepackage{times}
\usepackage{algorithm}
\usepackage{algorithmic}
\usepackage{graphicx} 
\usepackage{subfigure} 
\usepackage{microtype}
\usepackage[sort&compress]{natbib}
\usepackage{algorithm}
\usepackage{algorithmic}
\usepackage{amsmath}
\usepackage{amssymb}
\usepackage{flushend}
\usepackage{booktabs}
\usepackage{multirow}
\usepackage{booktabs}
\usepackage{todonotes}

\usepackage[accepted]{icml2014ed}

\usepackage{url}

\begin{document}\onecolumn
\icmltitle{Convolutional LSTM Networks for Subcellular Localization of Proteins}
\icmlauthor{S\o ren Kaae S\o nderby$^\text{1*}$}{skaaesonderby@gmail.com}
\icmlauthor{Casper Kaae S\o nderby$^\text{1*}$}{casperkaae@gmail.com}
\icmlauthor{Henrik Nielsen$^\text{3}$}{hnielsen@cbs.dtu.dk}
\icmlauthor{Ole Winther$^\text{1,2}$}{olwi@dtu.dk}
\icmladdress{
$^\text{1}$ Bioinformatics Centre, Department of Biology, University of Copenhagen, Copenhagen, Denmark\\ 
$^\text{2}$ Department for Applied Mathematics and Computer Science, Technical University of Denmark, 2800 Lyngby, Denmark\\
$^\text{3}$ Center for Biological Sequence Analysis, Department of Systems Biology, Technical University of Denmark, 2800 Lyngby, Denmark \\
$^\text{*}$ These authors contributed equally to the work}

\icmlkeywords{Subcellular location, machine learning, LSTM, RNN, neural networks, deep learning, convolution}
\icmltitlerunning{Convolutional LSTM Networks for Subcellular Localization of Proteins}
\vskip 0.3in

\begin{abstract}
Machine learning is widely used to analyze biological sequence data. Non-sequential models such as SVMs or feed-forward neural networks are often used although they have no natural way of handling sequences of varying length. Recurrent neural networks such as the long short term memory (LSTM) model on the other hand are designed to handle sequences. In this study we demonstrate that LSTM networks predict the subcellular location of proteins given only the protein sequence with high accuracy (0.902) outperforming current state of the art algorithms. We further improve the performance by introducing convolutional filters and experiment with an attention mechanism which lets the LSTM focus on specific parts of the protein. Lastly we introduce new visualizations of both the convolutional filters and the attention mechanisms and show how they can be used to extract biological relevant knowledge from the LSTM networks.
\end{abstract}

\setcounter{footnote}{0}
\section{INTRODUCTION} 
\begin{figure}[htbp]
	\centering
  	\includegraphics[width=1\textwidth]{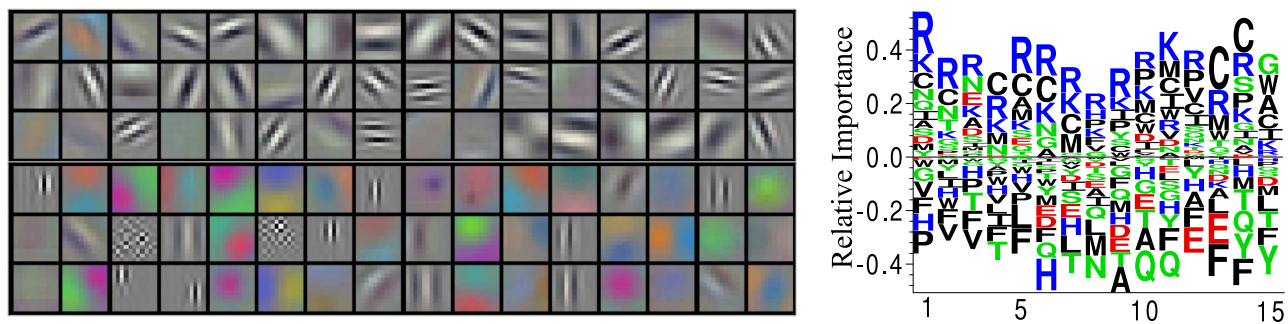}
	\caption{\textit{Left:} First layer convolutional filters learned in \cite{Krizhevsky2012}, note that 
many filters are edge detectors or color detectors. \textit{Right:} Example of learned filter on amino acid sequence data, note that this filter is sensitive to positively charged amino acids.}
	\label{fig:Krizhevsky}
\end{figure}

 Deep neural networks have gained popularity for a wide range of classification tasks in image recognition and speech tagging \cite{Dahl2012,Krizhevsky2012} and recently also within biology for prediction of exon skipping events \cite{Xiong2014}. Furthermore a surge of interest in recurrent neural networks (RNN) has followed the recent impressive results shown on challenging sequential problems like machine translation and speech recognition \cite{Graves2014,Sutskever2014,Bahdanau2014}. %
Within biology, sequence analysis is a very common task used for prediction of features in protein or nucleic acid sequences. Current methods generally rely on neural networks and support vector machines (SVM), which have no natural way of handling sequences of varying length. Furthermore these systems rely on highly hand-engineered input features requiring a high degree of domain knowledge when designing the algorithms \cite{Petersen2011,Emanuelsson2007}. This paper uses the long short term memory network (LSTM) \cite{Hochreiter1997} to analyze biological sequences and predict to which subcellular compartment a protein belongs. This prediction task, known as protein sorting or subcellular localization, has attracted large interest in the bioinformatics field \cite{Emanuelsson2007}. We show that an LSTM network, using only the protein sequence information, has significantly better performance than current state of the art SVMs and furthermore have nearly as good performance as large hand engineered systems relying on extensive metadata such as GO terms and evolutionary phylogeny, see Figure \ref{fig:lstm_model} \cite{Hoglund2006,Blum2009,Briesemeister2009}. These results show that LSTM networks are efficient algorithms that can be trained even on relatively small datasets of around 6000 protein sequences. %
Secondly we investigate how an LSTM network recognizes the sequence. In image recognition, convolutional neural networks (CNN) have shown state of the art performance in several different tasks \cite{Cun1990,Krizhevsky2012}. Here the lower layers of a CNN can often be interpreted as feature detectors recognizing simple geometric entities, see Figure \ref{fig:Krizhevsky}. We develop a simple visualization technique for convolutional filters trained on either DNA or amino acid sequences and show that in the biological setting filters can be interpreted as motif detectors, as visualized in Figure \ref{fig:Krizhevsky}. %
Thirdly, inspired by the work of Bahdanau et al., we augment the LSTM network with an attention mechanism that learns to assign importance to specific parts of the protein sequence. Using the attention mechanism we can visualize where the LSTM assigns importance, and we show that the network focuses on regions that are biologically plausible. Lastly we show that the LSTM network learns a fixed length representation of amino acids sequences that, when visualized, separates the sequences into clusters with biological meaning. %
The contributions of this paper are: %
\begin{enumerate}
\setlength\itemsep{-0.3em}
\item We show that LSTM networks combined with convolutions are efficient for predicting subcellular localization of proteins from sequence. %
\item We show that convolutional filters can be used for amino acid sequence analysis and introduce a visualization technique.%
\item We investigate an attention mechanism that lets us visualize where the LSTM network focuses.%
\item We show that the LSTM network effectively extracts a fixed length representation of variable length proteins.%
\end{enumerate}%
%
\section{MATERIALS AND METHODS}%
\subsection{MODEL}
This section introduces the LSTM cell and then explains how a regular LSTM (R-LSTM) can produce a single output. We then introduce the LSTM with attention mechanism (A-LSTM), and describes how the attention mechanism is implemented.%
\subsubsection{LSTM NETWORK}%
\begin{figure}[htbp]
    \centering
    \begin{minipage}{0.45\textwidth}
	\begin{center}
		\includegraphics[width=1\textwidth]{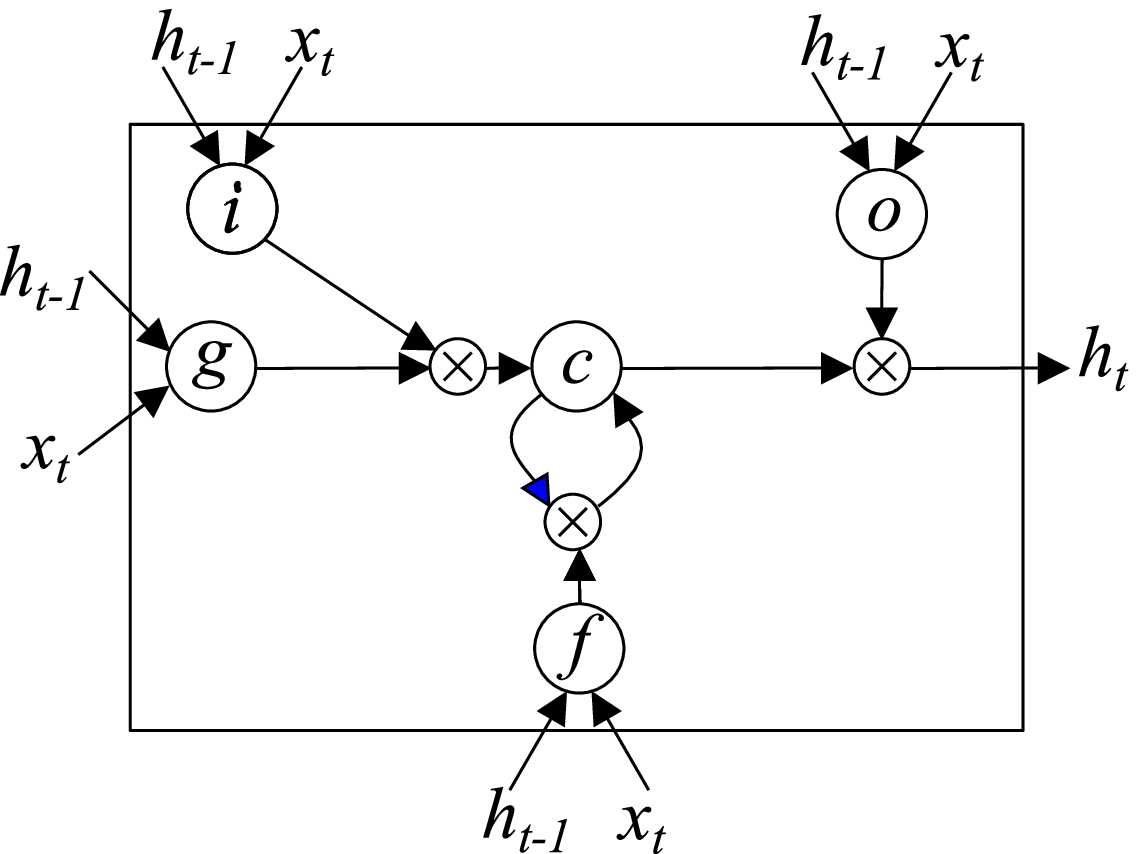}
	\end{center}
	\caption{LSTM memory cell. 
	         \textit{i}: input gate, 
	         \textit{f}: forget gate, 
	         \textit{o}: output gate, 
	         \textit{g}: input modulation gate, 
	         \textit{c}: memory cell. The Blue arrow heads refers to $c_{t-1}$.
	         The notation corresponds to equations \ref{eq:it} to \ref{eq:xt} such that $W_{xo}$ denotes wights for $x$ to output gate
	         and $W_{hf}$ denotes weights for $h_{t-1}$ to forget gates etc. Adapted from \cite{Zaremba2014a}.}
	\label{fig:cell}
    \end{minipage}$\quad$
    \begin{minipage}{0.45\textwidth}
	\begin{center}
		\includegraphics[width=1\textwidth]{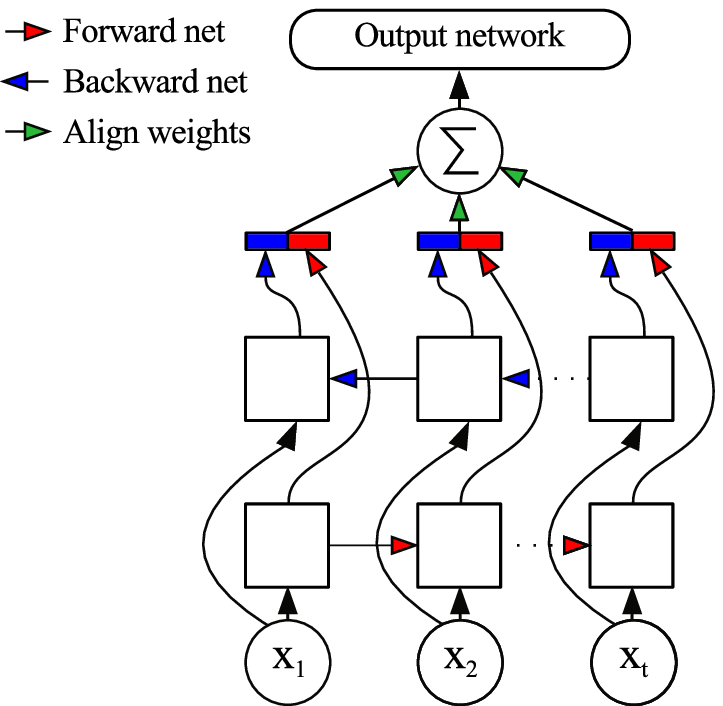}
	\end{center}\vspace{0.9cm}
	\caption{A-LSTM network. Each state of the hidden units, $h_t$ are weighted and summed before the output network calculates the predictions.}
	\label{fig:lstm_targets}
    \end{minipage}
\end{figure}

The LSTM cell is implemented as described in \cite{Graves2013} except for peepholes, because recent papers have shown good performance without \cite{Zaremba2014a,Zaremba2014b,Sutskever2014}. Figure \ref{fig:cell} shows the LSTM cell. Equations (\ref{eq:it})-(\ref{eq:xt}) state the forward recursions for a single LSTM layer. 
\begin{align}
i_t  &= \sigma(D(x_{t})W_{xi} + h_{t-1}W_{hi}  + b_{i})  \label{eq:it}\\	
f_t  &= \sigma(D(x_{t})W_{xf} + h_{t-1}W_{hf}  + b_{f}) \\
g_t  &= \tanh(D(x_{t}W_{xg}) + h_{t-1}W_{hg} + b_{g}) \\
c_t  &= f_t \odot c_{t-1} +  i_t \odot g_t  \\
o_t  &= \sigma(D(x_{t})W_{xo} + h_{t-1}W_{ho}  + b_{o}) \\
h_t  &= o_t \odot \tanh(c_{t})  \label{eq:ht}\\
\sigma(z)    &= \frac{1}{1+\exp(-z)} \\
\odot &: \text{Elementwise multiplication} \\
D     &: \text{Dropout, set values to zero with probability } p \\
x_t  &: \text{input from the previous layer: } h_t^{l-1}  \label{eq:xt}
\end{align}%
Where all quantities are given as row-vectors and activation and dropout functions are applied element-wise. If dropout is used it is only applied to non-recurrent connections in the LSTM cell \cite{Zaremba2014}. In a multilayer LSTM $h_t$ is passed upwards to the next layer.

\begin{figure}[tb]
	\begin{center}
		\includegraphics[width=1\textwidth]{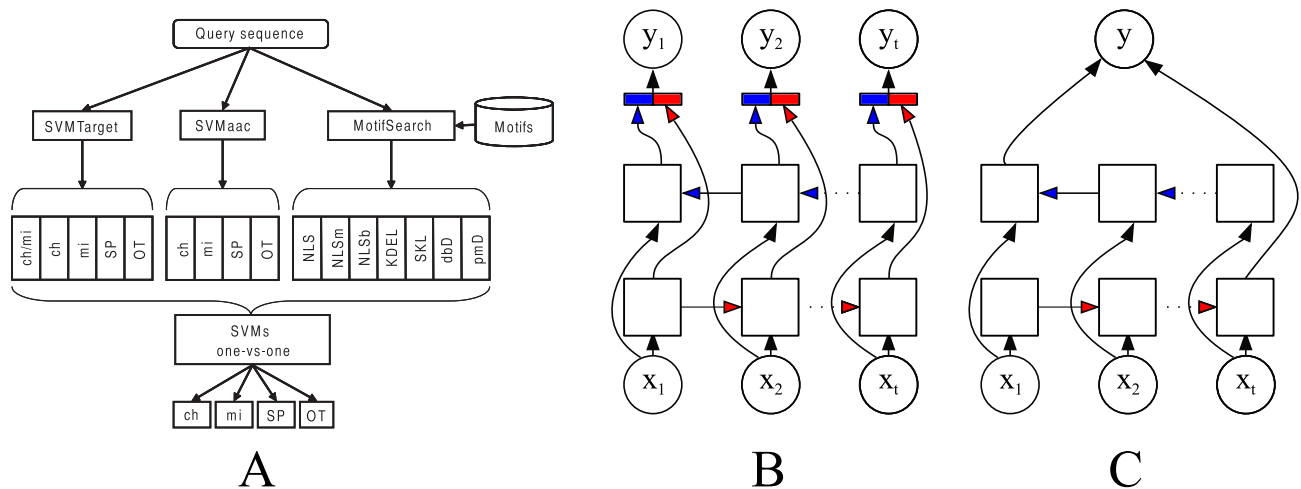}
	\end{center}
	\caption{\textit{A:} Schematic indicating how MultiLoc combines predictions from several sources to make predictions whereas the LSTM networks only rely on the sequence \cite{Hoglund2006}. \textit{B:} Unrolled single layer BLSTM. The forwards LSTM (red arrows) starts at time $1$ and the backwards LSTM (blue arrows) starts at time $T$, then they go forwards and backwards respectively. The errors from the forward and backward nets are combined and a prediction is made for each sequence position. Adapted from \cite{Graves2012}. \textit{C:} Unidirectional LSTM for predicting a single target. All targets except for the target at the last position are masked. Squares are LSTM layers.}
	\label{fig:lstm_model}
\end{figure}%

\subsection{REGULAR LSTM NETWORKS FOR PREDICTING SINGLE TARGETS}
When used for predicting a single target for each input sequence, one approach is to output the predicted target from the LSTM network at the last sequence position as shown in Figure \ref{fig:lstm_model}. A problem with this approach is that the gradient has to flow from the last position to all previous positions and that the LSTM network has to store information about all previously seen data in the last hidden state. Furthermore a regular bidirectional LSTM (BLSTM)\cite{Schuster1997} is not useful in this setting because the backward LSTM will only have seen a single position, $x_T$, when the prediction has to be made. We instead combine two unidirectional LSTMs, as shown in figure \ref{fig:lstm_model}C, where the backward LSTM has the input reversed. The prediction from the two LSTMs are combined before predictions.%
\subsection{ATTENTION MECHANISM LSTM NETWORK}%
Bahdanau et al. \cite{Bahdanau2014}, have introduced an attention mechanism for combining hidden state information from a encoder-decoder RNN approach to machine translation. The novelty in their approach is that they use an alignment function that for each output word finds important input words, thus aligning and translating at the same time. We modify this alignment procedure such that only a single target is produced for each sequence. The developed attention mechanism can be seen as assigning importance to each position in the sequence with respect to the prediction task. We use a BLSTM to produce a hidden state at each position and then use an attention function to assign importance to each hidden state as illustrated in figure \ref{fig:lstm_targets}. The weighted sum of hidden states is used as a single representation of the entire sequence. This modification allows the BLSTM model to naturally handle tasks involving prediction of a single target per sequence. Conceptually this corresponds to adding weighted skip connections (green arrow heads Figure \ref{fig:lstm_targets}) between any $h_t$ and the output network, with the weight on each skip connection being determined by the attention function.
Each hidden state $h_t$, $t=1,\ldots,T$ is used as input to a Feed Forward Neural Network (FFN) attention function:
\begin{equation}
	a_t = \tanh(h_t W_a) v^{\rm T}_a \  , 
\end{equation}
where $W_a$ is an attention hidden weight matrix and $v_a$ is an attention output vector. From the attention function we form softmax weights:
\begin{align}
\alpha_t  &=  \frac{\exp(a_t)}{{\Sigma^{T}_{t'=1} \exp{(a_{t'}} })}  \label{eq:alpha}
\end{align}%
that are used to produce a context vector $c$ as a convex combination of $T$ hidden states:
\begin{align}
c & = \Sigma^T_{t=1} h_t \alpha_t \ . \label{eq:c}
\end{align}%
The context vector is then used is as input to the classification FFN $f(c)$. 
We define $f$ as a single layer FFN with softmax outputs.
\subsection{SUBCELLULAR LOCALIZATION DATA}
The model was trained and evaluated on the dataset used to train the MultiLoc algorithm published by  Höglund et al. \cite{Hoglund2006}\footnote{\url{http://abi.inf.uni-tuebingen.de/Services/MultiLoc/multiloc_dataset}}. The dataset contains 5959 proteins annotated to one of 11 different subcellular locations. To reduce computational time the protein sequences were truncated to length 1000. We truncated by removing from the middle of the protein as both the N- and C-terminal regions are known to contain sorting signals \cite{Emanuelsson2007}. Each amino acid was encoded using 1-of-K encoding, the BLOSUM80 \cite{henikoff1992} and HSDM \cite{Prlic2000} substitution matrices and sequence profiles, yielding 80 features per amino acids. Sequence profiles where created with ProfilePro\footnote{\url{http://download.igb.uci.edu/}} using 3 blastpgp\footnote{\url{http://nebc.nox.ac.uk/bioinformatics/docs/blastpgp.html}} iterations on UNIREF50 \cite{magrane2011uniprot}. %
\subsection{VISUALIZATIONS}
Convolutional filters for images can be visualized by plotting the convolutional weights as pixel intensities as shown in figure \ref{fig:Krizhevsky}. However a similar approach does not make sense for amino acid inputs due to the 1-of-K vector encoding. Instead we view the 1D convolutions as a position specific scoring matrix (PSSM). The convolutional weights can be reshaped into a matrix of $l_{filter}$-by-$l_{enc}$, where the amino acid encoding length is is 20. Because the filters show relative importance we rescale all filters such that the height of the highest column is 1. Each filter can then be visualized as a PSSM logo, where the height of each column can be interpreted as position importance and the height of each letter is amino acid importance. We use Seq2logo with the PSSM-logo setting to create the convolution filter logos \cite{Thomsen2012}. \\\\%
We visualize the importance the A-LSTM network assigns to each position in the input by plotting $\alpha$ from equation \ref{eq:alpha}.  
Lastly we extract and plot the hidden representation from the LSTM networks. For the A-LSTM network we use $c$ from equation \ref{eq:c} and for the R-LSTM we use the last hidden state, $h_t$. Both $c$ and $h_t$ can be seen as fixed length representation of the amino acid sequences. We plot the representation using t-SNE \cite{VanDerMaaten2008}. 
\subsection{EXPERIMENTAL SETUP}%
All models were implemented in Theano \cite{Bastien2012} using a modified version of the Lasagne library\footnote{\url{https://github.com/skaae/nntools}} and trained with gradient descent. The learning rate was controlled with ADAM ($\alpha = 0.0002$, $\beta_1 = 0.1$, $\beta_2 = 0.001$, $\epsilon = 10^{−8}$ and $\lambda = 10^{-8}$) \cite{Kingma2014a}. Initial weights were sampled uniformly from the interval [-0.05, 0.05]. The network architecture is a 1D convolutional layer followed by an LSTM layer, a fully connected layer and a final softmax layer.
All layers use 50\% dropout. The 1D convolutional layer uses convolutions of sizes 1, 3, 5, 9, 15 and 21 with 10 filters of each size. Dense and convolutional layers use ReLU activation \cite{nair2010rectified} and the LSTM layer uses hyperbolic tangent. For the A-LSTM model the size of the first dimension of $W_a$ was 400. Based on previous experiments we trained for 100 epochs for all models and used 4/5 of the data for training the last 1/5 of the data for testing.

\section{RESULTS} 
\begin{figure}[htbp]
	\begin{center}
		\includegraphics[width=0.9\textwidth]{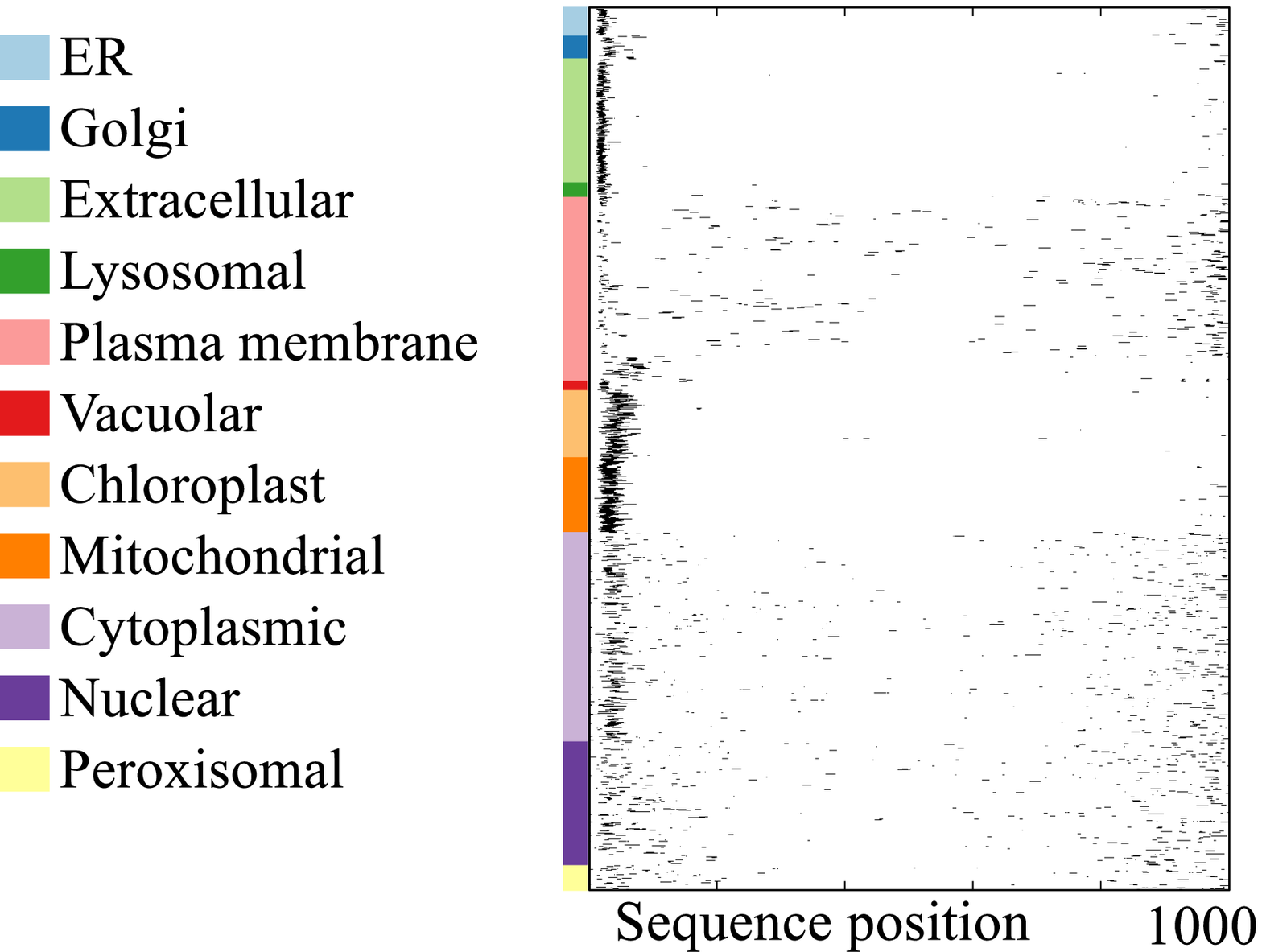}
	\end{center}
	\caption{Importance weights assigned to different regions of the proteins when making predictions. \textit{y}-axis is true group and \textit{x}-axis is the sequence positions. All proteins shorter than 1000 are zero padded from the middle such that the N and C
	terminals align.}
	\label{fig:aweights}
\end{figure}

\begin{figure}[htbp]
	\begin{center}
		\includegraphics[width=0.90\textwidth]{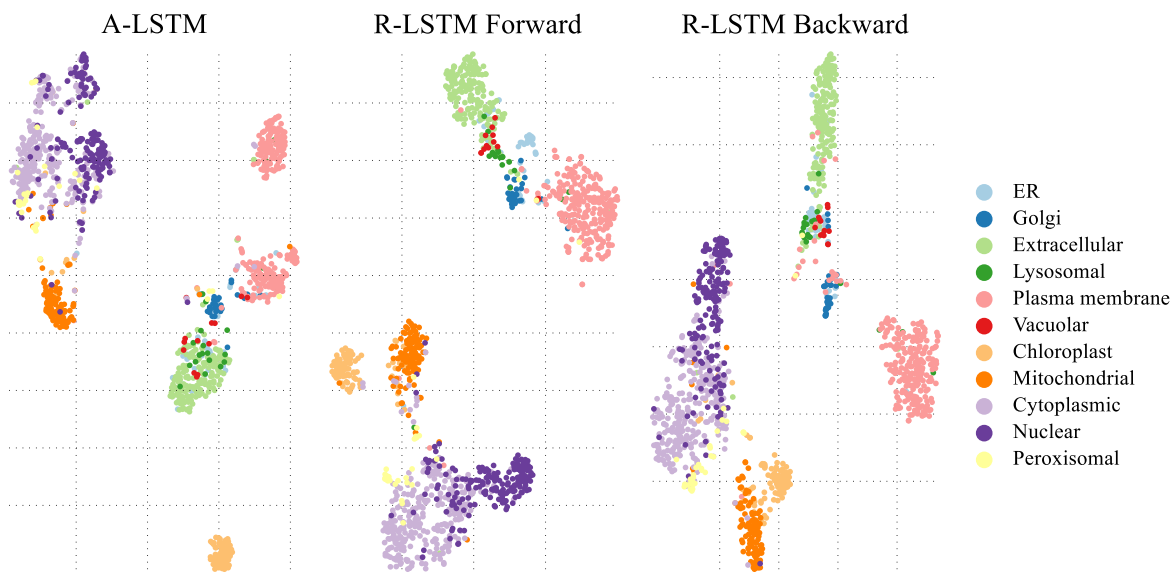}
	\end{center}
	\caption{t-SNE plot of hidden representation for Forward and Backward R-LSTM and A-LSTM.}
	\label{fig:tsne}
\end{figure}

\begin{figure}[ht]
	\begin{center}
		\includegraphics[width=0.75\textwidth]{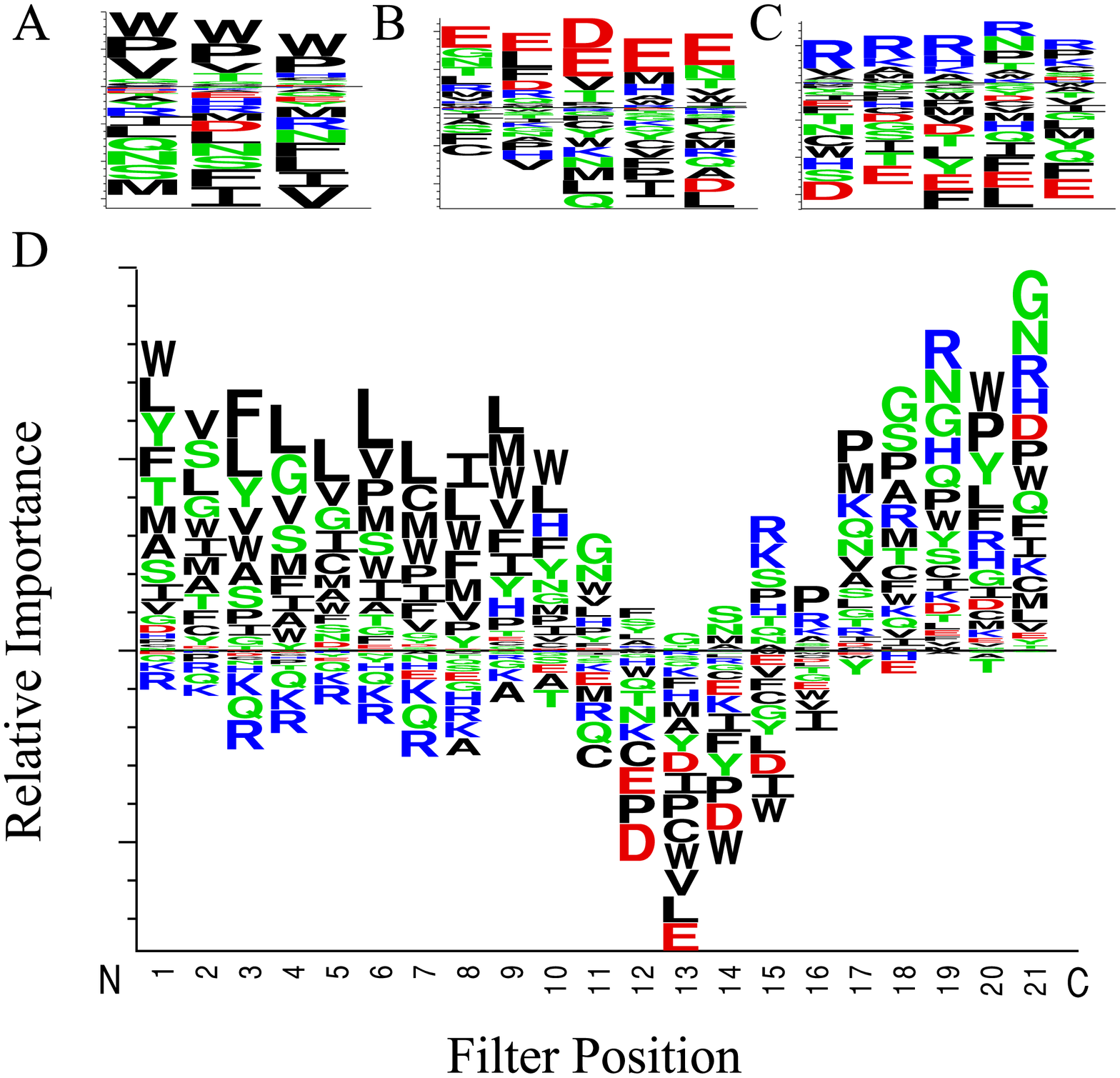}
	\end{center}
	\caption{Examples of learned filters. Filter A captures proline or trypthopan stretches, B) and C) are sensitive to positively and negatively charged regions, respectively. Note that for C, negative amino acids seems to suppress the output. Lastly we show a long filter which captures larger sequence motifs in the proteins.}
	\label{fig:filter1}
\end{figure}

Table \ref{tab:results} shows accuracy for the R-LSTM and A-LSTM models and several other models trained on the same dataset. Comparing the performance of the R-LSTM, A-LSTM and MultiLoc models, utilizing only the sequence information, the R-LSTM model (0.879 Acc.) performs better than the A-LSTM model (0.854 Acc.) whereas the MultiLoc model (0.767 Acc.) performs significantly worse. Furthermore the 10-ensemble R-LSTM model further increases the performance to 0.902 Acc. Comparing this performance with the other models, combining the sequence predictions from the MultiLoc model with large amounts of metadata for the final predictions, only the Sherloc2 model (0.930 Acc.) performs better than the R-LSTM ensemble.
Figure \ref{fig:aweights} shows a plot of the attention matrix from the A-LSTM model.
Figure \ref{fig:filter1} shows examples of the learned convolutional filters. Figure \ref{fig:tsne} shows the hidden state of the R-LSTM and the A-LSTM model. 
\begin{table}
\begin{minipage}{0.5\linewidth}
\caption{Comparison of results for LSTM models and MultiLoc1/2. MultiLoc1/2 accuracies are reprinted
from \cite{Goldberg2012} and the SherLoc accuracy from \cite{Briesemeister2009}.}
\label{tab:results}
\centering
\tiny
\begin{tabular}{@{}lll@{}}
\toprule
\bf{Model}         & \bf{Accuracy} &  \\ 
& & \\
\bf{Input: Protein Sequence} & & \\ \midrule
R-LSTM          & 0.879       &  \\
A-LSTM      & 0.854       &  \\
R-LSTM ensemble & \bf{0.902}       &  \\
MultiLoc      & 0.767        &  \\
& & \\
\bf{Input: Protein Sequence + Metadata} & & \\
\midrule
MultiLoc + PhyloLoc    & 0.842        &  \\
MultiLoc + PhyloLoc + GOLoc       & 0.871        &  \\
MultiLoc2     & 0.887        &  \\ 
SherLoc2      & \bf{0.930}        &  \\
\bottomrule
\end{tabular}
\end{minipage}$\quad$
\begin{minipage}{0.5\linewidth}
\caption{True labels are shown by row and model predictions by column. E.g. row 4 column 3 means that the actual class was Cytoplasmic but the model predicted Chloroplast.}
\label{tab:confusion}\centering\tiny
\begin{tabular}{@{}llllllllllll@{}}
\toprule
\textbf{Confusion Matrix} & & & & & & & & & & & \\
\midrule
\textbf{ER} 			 &26&1&0&0&8&1&0&0&0&3&0 \\
\textbf{Golgi} 			 &1&28&0&0&0&0&0&0&0&1&0\\
\textbf{Chloroplast}	 &0&0&82&3&0&0&5&0&0&0&0\\
\textbf{Cytoplasmic}     &0&0&1&266&0&0&3&12&0&0&0\\
\textbf{Extracellular}   &0&0&0&1&166&0&0&0&0&1&0\\
\textbf{Lysosomal}       &0&0&0&0&5&12&0&0&0&3&0\\
\textbf{Mitochondrial}   &0&0&2&5&0&0&94&1&0&0&0\\
\textbf{Nuclear}         &0&0&0&27&1&0&3&137&0&0&0\\
\textbf{Peroxisomal}     &0&1&0&10&0&0&0&1&18&2&0\\
\textbf{Plasma membrane} &0&0&0&0&5&0&1&1&0&241&0\\
\textbf{Vacuolar}        &0&0&0&0&7&0&0&0&0&1&5\\ \bottomrule
\end{tabular}
\end{minipage}%

\end{table}

\section{DISCUSSION AND CONCLUSION} 
In this paper we have introduced LSTM networks with convolutions for prediction of subcellular localization. Table \ref{tab:results} shows that the LSTM networks perform much better than other methods that only rely on information from the sequence (LSTM ensemble 0.902 vs. MultiLoc 0.767). This difference is all the more remarkable given that our method is biologically na\"{i}ve, only utilizing the sequences and their localization labels, while MultiLoc incorporates specific domain knowledge such as known motifs and signal anchors. One explanation for the performance difference is that the LSTM networks are able to look at both global and local sequence features whereas the SVM based models do not model global dependencies. The LSTM networks have nearly as good performance as methods that use information obtained from other sources than the sequence (LSTM ensemble 0.902 vs. SherLoc2 0.930). Incorporating these informations into the LSTM models could further improve the performance of these models. However, it is our opinion that using sequence alone yields the biologically most relevant prediction, while the incorporation of, e.g., GO terms limits the usability of the prediction requiring similar proteins to be already annotated to some degree. Furthermore, as we show below, a sequence-based method potentially allows for a de novo identification of sequence features essential for biological function.\\\\%

%
%
Figure \ref{fig:aweights} shows where in the sequence the A-LSTM network assigns importance. Sequences from the compartments ER, extracellular, lysosomal, and vacuolar all belong to the secretory pathway and contain N-terminal signal peptides, which are clearly seen as bars close to the left edge of the plot. Some of the ER proteins additionally have bars close to the right edge of the plot, presumably representing KDEL-type retention signals.  Golgi proteins are special in this context, since they are type II transmembrane proteins with signal anchors, slightly further from the N-terminus than signal peptides \cite{Hoglund2006}. Chloroplast and mitochondrial proteins also have N-terminal sorting signals, and it is apparent from the plot that chloroplast transit peptides are longer than mitochondrial transit peptides, which in turn are longer than signal peptides \cite{Emanuelsson2007}. For the plasma membrane category we see that some proteins have signal peptides, while the model generally focuses on signals, presumably transmembrane helices, scattered across the rest of the sequence with some overabundance close to the C-terminus. Some of the attention focused near the C-terminus could also represent signals for glycosylphosphatidylinositol (GPI) anchors \cite{Emanuelsson2007}. Cytoplasmic and nuclear proteins do not have N-terminal sorting signals, and we  see that the attention is scattered over a broader region of the sequences. However, especially for the cytoplasmic proteins there is some attention focused close to the N-terminus, presumably in order to check for the absence of signal peptides. Finally, peroxisomal proteins are known to have either N-terminal or C-terminal sorting signals (PTS1 and PTS2) \cite{Emanuelsson2007}, but these do not seem to have been picked up by the attention mechanism. \\\\%
In Figure \ref{fig:filter1} we investigate what the convolutional filters in the model focus on. Notably the short filters focus on amino acids with specific characteristics, such as positively or negatively charged, whereas the longer filters seem to focus on distributions of amino acids across longer sequences. The arginine-rich motif in Figure 7C could represent part of a nuclear localization signal (NLS), while the longer motif in Figure 7D could represent the transition from transmembrane helix (hydrophobic) to cytoplasmic loop (in accordance with the "positive-inside" rule). We believe that the learned filters can be used to discover new sequence motifs for a large range of protein and genomic features. \\\\%
In Figure \ref{fig:tsne} we investigate whether the LSTM models are able to extract fixed length representations of variable length proteins. Using t-SNE we plot the LSTMs hidden representation of the sequences. It is apparent that proteins from the same compartment generally group together, while the cytoplasmic and nuclear categories tend to overlap. The corresponds with the fact that these two categories are relatively often confused, see Table \ref{tab:confusion}. The categories form clusters which make biological sense; all the proteins with signal peptides (ER, extracellular, lysosomal, and vacuolar) lie close to each other in t-SNE space in all three plots, while the proteins with other N-terminal sorting signals (chloroplasts and mitochondria) are close in the R-LSTM plots (but not in the A-LSTM plot). Note that the lysosomal and vacuolar categories are very close to each other in the plots, this corresponds with the fact that these two compartments are considered homologous \cite{Hoglund2006}. \\\\%
In summary we have introduced LSTM networks with convolutions for subcellular
localization. By visualizing the learned filters we have shown that these
can be interpreted as motif detectors, and lastly we have shown that the LSTM
network can represent protein sequences as a fixed length vector in a 
representation that is biologically interpretable. %

\begin{footnotesize}
\bibliographystyle{icml2015}
\bibliography{library.bib}
\end{footnotesize}

\end{document}